\documentclass[a4paper,12pt]{article}
\usepackage[left=3.17cm,right=3.17cm,top=2.54cm,
headheight=0.5cm,headsep=0.54cm,bottom=2.54cm,footskip=0.79cm
]{geometry}

\usepackage{amsmath,amssymb,dsfont}

\usepackage{hyperref}

\begin{document}

\title{Connes distance of $2D$ harmonic oscillators in quantum phase space}

\author{Bing-Sheng Lin$^{1,2,\dag}$,\,\,Tai-Hua Heng$^{\,3,\ddagger}$\\
\small $^{1}$School of Mathematics, South China University of Technology,\\
\small Guangzhou 510641, China\\
\small$^{2}$Laboratory of Quantum Science and Engineering,\\
\small South China University of Technology, Guangzhou 510641, China\\
\small $^{\dag}$Email: sclbs@scut.edu.cn\\
\small $^{3}$School of Physics and Material Science, Anhui University, Hefei 230601, China\\
\small $^{\ddagger}$Email: hength@ahu.edu.cn
}

\date{}

\maketitle

\begin{abstract}
We study the Connes distance of quantum states of $2D$ harmonic oscillators in phase space. Using the Hilbert-Schmidt operatorial formulation, we construct a boson Fock space and a quantum Hilbert space, and obtain the Dirac operator and a spectral triple corresponding to a $4D$ quantum phase space. Based on the ball condition, we obtain some constraint relations about the optimal elements. We construct the explicit expressions of the corresponding optimal elements and then derive the Connes distance between two arbitrary Fock states of $2D$ quantum harmonic oscillators. We prove that these two-dimensional distances satisfy the Pythagoras theorem.
\\

\end{abstract}

\section{Introduction}
In quantum physics, the physical quantities are represented by some operators in a Hilbert space. Usually, these operators are noncommutative with each other.
So it is natural to study physical systems by noncommutative geometry.
In the 1980's, Connes formulated the mathematically rigorous framework of noncommutative geometry \cite{Connes}.
Many kinds of noncommutative spaces are studied by physicists and mathematicians, such as the Moyal plane and fuzzy space \cite{Moyal}-\cite{Grosse}.
The phase space in quantum mechanics is also a Moyal-type noncommutative space, because the position and momentum operators satisfy the noncommutative Heisenberg algebras.

Due to the noncommutativity, there is no traditional point in a noncommutative space. This is different from normal commutative spaces. We only have states in noncommutative spaces. So there is no normal distance between two points in a noncommutative space, but we can calculate some kinds of distance measures between the states, such as the Connes distance \cite{Connes1}.
Many researchers have already studied the Connes distance in some kinds of noncommutative spaces.
Bimonte \emph{et. al.} calculated the distances between the points of a lattice on which the usual discretized Dirac operator has been defined \cite{Bimonte}.
Cagnache \emph{et. al.} have studied the Connes spectral distance in the Moyal plane \cite{Cagnache}. They explicitly computed Connes spectral distance between the pure states which corresponding to eigenfunctions of the quantum harmonic oscillators.
Martinetti \emph{et. al.} have studied the metric aspect of the Moyal plane from Connes' noncommutative geometry point of view \cite{Martinetti}. They obtained the spectral distance between coherent states of the quantum harmonic oscillator as the Euclidean distance on the plane. They also computed the spectral distance in the so-called double Moyal plane.
D'Andrea \emph{et. al.} have studied the Pythagoras' theorem in noncommutative geometry \cite{DAndrea}. They showed that for non-pure states it is replaced by some Pythagoras inequalities.
Franco and Wallet also studied metrics and causality on Moyal planes \cite{Franco}.
Scholtz and his collaborators have done many works on the studies of Connes spectral distances in Moyal plane and fuzzy sphere \cite{Scholtz}-\cite{Revisiting}. They developed the Hilbert-Schmidt operatorial formulation, and obtained the distances of harmonic oscillator states and also coherent states.
Kumar \emph{et. al.} used Dirac eigen-spinor method to compute spectral distances in doubled Moyal plane \cite{Kumar}.
Chakraborty \emph{et. al.} also studied the spectral distance on Lorentzian Moyal plane \cite{Chakraborty}.

In the present work, we consider the Connes distance of the quantum states (namely the Fock states) of $2D$ harmonic oscillators. Actually, some researchers have already studied the Connes distance between the states of $1D$ harmonic oscillators \cite{Cagnache,Scholtz,Revisiting}. But we find that the calculations and results of the Connes distance of $2D$ harmonic oscillators are much more complicated than those of $1D$ oscillator system. So it is significant to study the explicit formulae of the Connes distance of $2D$ harmonic oscillator systems. The method used in the present work is also different from those used in the literatures.

This paper is organized as follows. In Sec.~\ref{sec2}, we consider the $4D$ quantum phase space and construct a corresponding spectral triple. Using the Hilbert-Schmidt operatorial formulation, we construct a boson Fock space and a quantum Hilbert space, and obtain the Dirac operator.
In Sec.~\ref{sec3}, we review the definition of Connes spectral distance, and derive some inequalities corresponding to the ball condition.
Based on these inequalities, we derive some constraint relations about the optimal elements in Sec.~\ref{sec4}.
The Connes distance between two adjacent Fock states of $2D$ harmonic oscillators is calculated in Sec.~\ref{sec5}.
In Sec.~\ref{sec6}, we also calculate the Connes distance between two arbitrary Fock states.
Some discussions and conclusions are given in Sec.~\ref{sec7}.

\section{4D quantum phase space and spectral triple}\label{sec2}
Let us consider the $4D$ quantum phase space, and the coordinate operators $\hat{x}_i$, $\hat{p}_i$ satisfy the following commutation relation,
\begin{equation}\label{qps}
[\hat{x}_i,\,\hat{p}_j]=\mathrm{i}\delta_{ij}\hbar\qquad(i,j=1,2),
\end{equation}
and others vanish.
We can define the following creation and annihilation operators,
\begin{equation}
\hat{a}_i=\frac{1}{\sqrt{2\hbar}}(\hat{x}_i+\mathrm{i}\hat{p}_i),\qquad
\hat{a}_i^{\dag}=\frac{1}{\sqrt{2\hbar}}(\hat{x}_i-\mathrm{i}\hat{p}_i),
\end{equation}
which satisfy the commutation relations $[\hat{a}_i,\hat{a}_j^{\dag}]=\delta_{ij}$, and $[\hat{a}_i,\hat{a}_j]=[\hat{a}_i^{\dag},\hat{a}_j^{\dag}]=0$.

In the present work, we will use the Hilbert-Schmidt operatorial formulation developed in Ref.~\cite{Formulation}. We can construct the following boson Fock space
\begin{equation}
\mathcal{F}=\mathrm{span}\left\{|n_1,n_2\rangle
=\frac{1}{\sqrt{n_1!n_2!}}(\hat{a}_1^{\dag})^{n_1}(\hat{a}_2^{\dag})^{n_2}|00\rangle,
\quad n_1,n_2=0,1,2,...\right\},
\end{equation}
where $|00\rangle$ is the vacuum state, $\hat{a}_i|00\rangle=0$. The Fock states $|n_1,n_2\rangle$ are eigenstates of the number operators $\hat{N}_i=\hat{a}_i^{\dag}\hat{a}_i$, $\hat{N}_i|n_1,n_2\rangle=n_i|n_1,n_2\rangle$, and
\begin{eqnarray}\label{a1a2}
&&\hat{a}_1|n_1,n_2\rangle=\sqrt{n_1}|n_1-1,n_2\rangle,\qquad \hat{a}_1^{\dag}|n_1,n_2\rangle=\sqrt{n_1+1}|n_1+1,n_2\rangle,\nonumber\\
&&\hat{a}_2|n_1,n_2\rangle=\sqrt{n_2}|n_1,n_2-1\rangle,\qquad \hat{a}_2^{\dag}|n_1,n_2\rangle=\sqrt{n_2+1}|n_1,n_2+1\rangle.
\end{eqnarray}
These Fock states can be regarded as the quantum states of a two-dimensional harmonic oscillator system.

Furthermore, one can construct the following quantum Hilbert space
\begin{equation}
\mathcal{Q}=\mathrm{span}\left\{|m_1,m_2\rangle\langle n_1,n_2|\right\}.
\end{equation}
In the followings, we will denote the elements $\psi(\hat{x}_i,\hat{p}_i)$ of the quantum Hilbert space by $|\psi)$ (The elements $\phi$ of the boson Fock space $\mathcal{F}$ are denoted by $|\phi\rangle$),
\begin{equation}
\psi(\hat{x}_i,\hat{p}_i)\equiv|\psi)=
\sum_{i,j,k,l}\psi^{i,j}_{k,l}|i,j\rangle\langle k,l|.
\end{equation}
The corresponding inner product is
\begin{eqnarray}
(\phi|\psi)
&=&\left(\phi(\hat{x}_i,\hat{p}_i),\psi(\hat{x}_i,\hat{p}_i)\right)
=\mathrm{tr}_\mathcal{F}\left(\phi(\hat{x}_i,\hat{p}_i)^{\dag}\psi(\hat{x}_i,\hat{p}_i)\right)\nonumber\\
&=&\sum_{m,n=0}^{\infty}\langle m,n|\phi(\hat{x}_i,\hat{p}_i)^{\dag}\psi(\hat{x}_i,\hat{p}_i)|m,n\rangle,
\end{eqnarray}
where $\mathrm{tr}_{\mathcal{F}}(\cdot)$ denotes the trace over $\mathcal{F}$.

The quantum phase space (\ref{qps}) is also a Moyal-type noncommutative space.
In general, a noncommutative space corresponds to a spectral triple $(\mathcal{A},\mathcal{H},\mathcal{D})$ with $\mathcal{A}$ an involutive algebra acting on a Hilbert space $\mathcal{H}$, and $\mathcal{D}$ is the Dirac operator on $\mathcal{H}$ \cite{Connes}.
Moyal spaces are non-compact spectral triples \cite{Gayral}.
One can construct a spectral triple $(\mathcal{A},\mathcal{H},\mathcal{D})$ for a $4D$ quantum phase space as follows,
\begin{equation}
\mathcal{A}=\mathcal{Q},
\qquad\mathcal{H}=\mathcal{F}\otimes \mathbb{C}^4,
\end{equation}
and an element $e\in \mathcal{A}$ acts on
$\Psi=\left(\begin{array}{c}
     |\psi_1\rangle \\
     |\psi_2\rangle \\
     |\psi_3\rangle \\
     |\psi_4\rangle \\
\end{array}\right)\in\mathcal{H}$
through the diagonal representation $\pi$ as
\begin{equation}
\pi(e)\Psi=\pi(e)\left(\begin{array}{c}
     |\psi_1\rangle \\
     |\psi_2\rangle \\
     |\psi_3\rangle \\
     |\psi_4\rangle \\
\end{array}\right)
=\left(\begin{array}{cccc}
e & 0 & 0 & 0 \\
0 & e & 0 & 0 \\
0 & 0 & e & 0 \\
0 & 0 & 0 & e \\
\end{array}\right)\left(\begin{array}{c}
     |\psi_1\rangle \\
     |\psi_2\rangle \\
     |\psi_3\rangle \\
     |\psi_4\rangle \\
\end{array}\right)
=\left(\begin{array}{c}
     e|\psi_1\rangle \\
     e|\psi_2\rangle \\
     e|\psi_3\rangle \\
     e|\psi_4\rangle \\
\end{array}\right).
\end{equation}

Similar to Ref.~\cite{Revisiting}, in order to construct the Dirac Operator for the $4D$ quantum phase space, one can consider the following extended noncommutative phase space in which the position operators $\hat{X}_i$, $\hat{Y}_i$ and the momentum operators $\hat{P}_i$, $\hat{Q}_i$ satisfy the following commutation relations \cite{Lin},
\begin{equation}\label{nc}
[\hat{X}_i,\,\hat{P}_j]=[\hat{Y}_i,\,\hat{Q}_j]=\mathrm{i}\delta_{ij}\hbar\,,\qquad
[\hat{X}_i,\,\hat{Y}_j]=\mathrm{i}\delta_{ij}\mu\,,\qquad
[\hat{P}_i,\,\hat{Q}_j]=\mathrm{i}\delta_{ij}\nu\,,
\end{equation}
and others vanish. Here $i,j=1,2$, and $\mu$, $\nu$ are some parameters.
A unitary representation of the noncommutative Heisenberg algebra (\ref{nc}) is obtained by the following actions on the quantum Hilbert space $\mathcal{Q}$:
\begin{eqnarray}
&\hat{X}_i|\phi)=|\hat{x}_i\phi),\qquad
&\hat{Y}_i|\phi)
=\frac{\mu}{\hbar}|\hat{p}_i\phi)
+\frac{\sqrt{\hbar^2-\mu\nu}}{\hbar}|\phi\hat{p}_i),
\nonumber\\
&\hat{P}_i|\phi)=|\hat{p}_i\phi),\qquad
&\hat{Q}_i|\phi)
=-\frac{\nu}{\hbar}|\hat{x}_i\phi)
+\frac{\sqrt{\hbar^2-\mu\nu}}{\hbar}|\phi\hat{x}_i).
\end{eqnarray}

By virtue of the result in Ref.~\cite{Gayral}, the Dirac operator $\mathcal{D}$ can be written as
\begin{equation}\label{do0}
\mathcal{D}
=\frac{1}{\mu}\gamma^1\hat{Y}_1+\frac{1}{\nu}\gamma^2\hat{Q}_1
+\frac{1}{\mu}\gamma^3\hat{Y}_2+\frac{1}{\nu}\gamma^4\hat{Q}_2,
\end{equation}
where $\gamma^k$'s are the Euclidean Dirac matrices satisfying
\begin{equation}
\gamma^k\gamma^l+\gamma^l\gamma^k=2\delta_{kl}\mathds{I}_4\qquad (k,l=1,2,3,4),
\end{equation}
and $\mathds{I}_4$ is the $4\times 4$ identity matrix. For example, one can choose
\begin{eqnarray}
&&\gamma^1=\left(
  \begin{array}{cccc}
    0 & 0 & 0 & \mathrm{i} \\
    0 & 0 & \mathrm{i} & 0 \\
    0 & -\mathrm{i} & 0 & 0 \\
    -\mathrm{i} & 0 & 0 & 0 \\
  \end{array}\right),
\qquad
\gamma^2=\left(
  \begin{array}{cccc}
    0 & 0 & 0 &1 \\
    0 & 0 & -1 & 0 \\
    0 & -1 & 0 & 0 \\
    1 & 0 & 0 & 0 \\
  \end{array}\right),
\nonumber\\
&&\gamma^3=\left(
  \begin{array}{cccc}
    0 & 0 & \mathrm{i} & 0 \\
    0 & 0 & 0 & -\mathrm{i} \\
    -\mathrm{i} & 0 & 0 & 0 \\
    0 & \mathrm{i} & 0 & 0 \\
  \end{array}\right),
\qquad
\gamma^4=\left(
  \begin{array}{cccc}
    0 & 0 & 1 & 0 \\
    0 & 0 & 0 & 1 \\
    1 & 0 & 0 & 0 \\
    0 & 1 & 0 & 0 \\
  \end{array}\right),
\end{eqnarray}
and the Dirac operator (\ref{do0}) can be expressed as
\begin{equation}
\mathcal{D}
=\left(
  \begin{array}{cccc}
    0 & 0 & \mathrm{i}\frac{1}{\mu}\hat{Y}_2+\frac{1}{\nu}\hat{Q}_2 & \mathrm{i}\frac{1}{\mu}\hat{Y}_1+\frac{1}{\nu}\hat{Q}_1 \\
    0 & 0 & \mathrm{i}\frac{1}{\mu}\hat{Y}_1-\frac{1}{\nu}\hat{Q}_1 & -\mathrm{i}\frac{1}{\mu}\hat{Y}_2+\frac{1}{\nu}\hat{Q}_2 \\
    -\mathrm{i}\frac{1}{\mu}\hat{Y}_2+\frac{1}{\nu}\hat{Q}_2 & -\mathrm{i}\frac{1}{\mu}\hat{Y}_1-\frac{1}{\nu}\hat{Q}_1 & 0 & 0 \\
    -\mathrm{i}\frac{1}{\mu}\hat{Y}_1+\frac{1}{\nu}\hat{Q}_1 & \mathrm{i}\frac{1}{\mu}\hat{Y}_2+\frac{1}{\nu}\hat{Q}_2 & 0 & 0 \\
  \end{array}\right).
\end{equation}

After some straightforward calculations, one can obtain the commutator $[\mathcal{D},\pi(e)]$ acting on an element $\Phi\in \mathcal{Q}\otimes \mathbb{C}^4$ as
\begin{eqnarray}
\lefteqn{[\mathcal{D},\pi(e)]\Phi
=[\mathcal{D},\pi(e)]
\left(
  \begin{array}{c}
   |\phi_1) \\
   |\phi_2) \\
   |\phi_3) \\
   |\phi_4) \\
  \end{array}
\right)}\nonumber\\
&&=\frac{1}{\hbar}\!\left(
  \begin{array}{cccc}
    \!0 & \!0 &\![-\hat{x}_2{+}\mathrm{i}\hat{p}_2,e] & \![-\hat{x}_1{+}\mathrm{i}\hat{p}_1,e] \\
    \!0 & \!0 &\![\hat{x}_1{+}\mathrm{i}\hat{p}_1,e] & \![-\hat{x}_2{-}\mathrm{i}\hat{p}_2,e] \\{}
    \![-\hat{x}_2{-}\mathrm{i}\hat{p}_2,e] & \![\hat{x}_1{-}\mathrm{i}\hat{p}_1,e] & \!0 & \!0 \\{}
    \![-\hat{x}_1{-}\mathrm{i}\hat{p}_1,e] & \![-\hat{x}_2{+}\mathrm{i}\hat{p}_2,e] & \!0 & \!0 \\
  \end{array}\!\right)
\!\!\left(
  \begin{array}{c}
   \!|\phi_1) \\
   \!|\phi_2) \\
   \!|\phi_3) \\
   \!|\phi_4) \\
  \end{array}
\!\right)\!.~~
\end{eqnarray}
Regarding $\Phi$ as a test function, one can identify the Dirac operator $\mathcal{D}$ as
\begin{eqnarray}
\mathcal{D}=\frac{1}{\hbar}\left(
  \begin{array}{cccc}
    0 & 0 &-\hat{x}_2+\mathrm{i}\hat{p}_2 & -\hat{x}_1+\mathrm{i}\hat{p}_1 \\
    0 & 0 &\hat{x}_1+\mathrm{i}\hat{p}_1 & -\hat{x}_2-\mathrm{i}\hat{p}_2 \\{}
    -\hat{x}_2-\mathrm{i}\hat{p}_2 & \hat{x}_1-\mathrm{i}\hat{p}_1 & 0 & 0 \\{}
    -\hat{x}_1-\mathrm{i}\hat{p}_1 & -\hat{x}_2+\mathrm{i}\hat{p}_2 & 0 & 0 \\
  \end{array}\right),
\end{eqnarray}
or
\begin{eqnarray}\label{do}
\mathcal{D}=\sqrt{\frac{2}{\hbar}}\left(
  \begin{array}{cccc}
    0 & 0 & -\hat{a}_2^\dag & -\hat{a}_1^\dag \\
    0 & 0 &  \hat{a}_1 & -\hat{a}_2 \\
    -\hat{a}_2 & \hat{a}_1^\dag & 0 & 0 \\
    -\hat{a}_1 & -\hat{a}_2^\dag & 0 & 0 \\
  \end{array}\right).
\end{eqnarray}

\section{Connes spectral distance and ball condition}\label{sec3}
The Connes spectral distance between two states $\omega$ and $\omega'$ is defined as \cite{Connes1}
\begin{equation}
d(\omega,\omega')\equiv \sup_{e\in B}|\omega(e)-\omega'(e)|,
\end{equation}
where
\begin{equation}\label{ball}
B=\left\{e\in \mathcal{A}:\big\|[\mathcal{D},\pi(e)]\big\|_{op}\leqslant 1\right\},
\end{equation}
and
\begin{equation}
\|A\|_{op}\equiv\sup_{\psi\in \mathcal{H}}\frac{\|A\psi\|}{\|\psi\|},\qquad
\|A\|^2\equiv (A,A)=\mathrm{tr}_{\mathcal{F}}(A^{\dag}A).
\end{equation}
The inequality in (\ref{ball}) is the so-called ball condition.

Similar to Ref.~\cite{Scholtz}, here we only consider the case where the quantum states $\omega$ are normal and bounded so that they are representable by density matrices $\rho$.
The action of the state $\omega$ on an element $e\in \mathcal{A}$ can be written as
\begin{equation}
\omega(e)=\mathrm{tr}_{\mathcal{F}}(\rho e).
\end{equation}
Suppose the quantum states $\omega$ and $\omega'$ correspond to the density matrices $\rho$ and $\rho'$, respectively. We have
\begin{equation}
d(\omega,\omega')=\sup_{e\in B}|\mathrm{tr}_{\mathcal{F}}(\rho e)-\mathrm{tr}_{\mathcal{F}}(\rho' e)|
=\sup_{e\in B}|\mathrm{tr}_{\mathcal{F}}(\Delta\rho\, e)|=\sup_{e\in B}|(\Delta\rho,e)|,
\end{equation}
where $\Delta\rho=\rho-\rho'$, and $(\Delta\rho)^{\dag}=\Delta\rho$.

Now let us calculate the commutator $[\mathcal{D},\pi(e)]$ for a Hermitian element $e\in \mathcal{A}$.
Using the Dirac operator $\mathcal{D}$ (\ref{do}), we have
\begin{eqnarray}
[\mathcal{D},\pi(e)]&=&\sqrt{\frac{2}{\hbar}}\left(
  \begin{array}{cccc}
    0 & 0 & [-\hat{a}_2^\dag,e] & [-\hat{a}_1^\dag,e] \\
    0 & 0 &  [\hat{a}_1,e] & [-\hat{a}_2,e] \\{}
    [-\hat{a}_2,e] & [\hat{a}_1^\dag,e] & 0 & 0 \\{}
    [-\hat{a}_1,e] & [-\hat{a}_2^\dag,e] & 0 & 0 \\
  \end{array}\right)\nonumber\\
&=&\sqrt{\frac{2}{\hbar}}\left(
  \begin{array}{cccc}
    0 & 0 & [\hat{a}_2,e]^\dag & [\hat{a}_1,e]^\dag \\
    0 & 0 &  [\hat{a}_1,e] & -[\hat{a}_2,e] \\{}
    -[\hat{a}_2,e] & -[\hat{a}_1,e]^\dag & 0 & 0 \\{}
    -[\hat{a}_1,e] & [\hat{a}_2,e]^\dag & 0 & 0 \\
  \end{array}\right)\nonumber\\
&=&\sqrt{\frac{2}{\hbar}}\left(
  \begin{array}{cc}
    0 & D_1 \\
    -D_1^\dag & 0 \\
  \end{array}\right),
\end{eqnarray}
where
\begin{equation}\label{d1d}
D_1=\left(
  \begin{array}{cc}
  [\hat{a}_2,e]^\dag & [\hat{a}_1,e]^\dag \\{}
  [\hat{a}_1,e] & -[\hat{a}_2,e] \\
  \end{array}\right),\quad
D_1^\dag=\left(
  \begin{array}{cc}
  [\hat{a}_2,e] & [\hat{a}_1,e]^\dag \\{}
  [\hat{a}_1,e] & -[\hat{a}_2,e]^\dag \\
  \end{array}\right).
\end{equation}

It is easy to verify that
\begin{equation}
\left\|[\mathcal{D},\pi(e)]^\dag[\mathcal{D},\pi(e)]\right\|_{op}
=\frac{2}{\hbar}\max\left\{\|D_1^\dag D_1\|_{op},\|D_1 D_1^\dag\|_{op}\right\}.
\end{equation}
It is known that for any bounded operator $A$, there is $\|A\|_{op}^2=\|A^\dag\|_{op}^2=\|A^\dag A\|_{op}$. Therefore we have
\begin{eqnarray}
\big\|[\mathcal{D},\pi(e)]\big\|_{op}^2
&=&\left\|[\mathcal{D},\pi(e)]^\dag[\mathcal{D},\pi(e)]\right\|_{op}\nonumber\\
&=&\frac{2}{\hbar}\|D_1^\dag D_1\|_{op}=\frac{2}{\hbar}\|D_1 D_1^\dag\|_{op}\nonumber\\
&=&\frac{2}{\hbar}\|D_1\|_{op}^2=\frac{2}{\hbar}\|D_1^\dag\|_{op}^2.
\end{eqnarray}
So for any Hermitian element $e\in B$, using the ball condition in (\ref{ball}), one can obtain the following inequality,
\begin{equation}\label{bc}
\|D_1^\dag D_1\|_{op}=\|D_1 D_1^\dag\|_{op}
=\frac{\hbar}{2}\big\|[\mathcal{D},\pi(e)]\big\|_{op}^2
\leqslant \frac{\hbar}{2}.
\end{equation}

From (\ref{d1d}), one can obtain
\begin{eqnarray}
D_1^\dag D_1&=&\left(
  \begin{array}{cc}
  [\hat{a}_2,e] & [\hat{a}_1,e]^\dag \\{}
  [\hat{a}_1,e] & -[\hat{a}_2,e]^\dag \\
  \end{array}\right)\left(
  \begin{array}{cc}
  [\hat{a}_2,e]^\dag & [\hat{a}_1,e]^\dag \\{}
  [\hat{a}_1,e] & -[\hat{a}_2,e] \\
  \end{array}\right)\nonumber\\
&=&\left(
  \begin{array}{cc}
  [\hat{a}_2,e][\hat{a}_2,e]^\dag+[\hat{a}_1,e]^\dag[\hat{a}_1,e] & [\hat{a}_2,e][\hat{a}_1,e]^\dag-[\hat{a}_1,e]^\dag[\hat{a}_2,e] \\{}
  [\hat{a}_1,e][\hat{a}_2,e]^\dag-[\hat{a}_2,e]^\dag[\hat{a}_1,e] & [\hat{a}_1,e][\hat{a}_1,e]^\dag+[\hat{a}_2,e]^\dag[\hat{a}_2,e] \\
  \end{array}\right).
\end{eqnarray}
Consider an operator $A$ with the matrix elements $A_{ij}$ in some orthonormal bases, there is the following Bessel's inequality \cite{Revisiting}
,
\begin{equation}
\|A\|_{op}^2\geqslant \sum_i|A_{ij}|^2\geqslant |A_{ij}|^2.
\end{equation}
So we have
\begin{equation}\label{d1d1}
\|D_1^\dag D_1\|_{op}
\geqslant\!\sup_{\phi\in\mathcal{F},\langle\phi|\phi\rangle=1}
\langle\phi|[\hat{a}_1,e]^\dag[\hat{a}_1,e]+[\hat{a}_2,e][\hat{a}_2,e]^\dag|\phi\rangle.
\end{equation}
From (\ref{bc}) and (\ref{d1d1}), for any Hermitian element $e\in B$, there is the following inequality
\begin{equation}\label{bc1}
\sup_{\phi\in\mathcal{F},\langle\phi|\phi\rangle=1}
\langle\phi|[\hat{a}_1,e]^\dag[\hat{a}_1,e]+[\hat{a}_2,e][\hat{a}_2,e]^\dag|\phi\rangle
\leqslant\frac{\hbar}{2}.
\end{equation}
Similarly, we have
\begin{equation}\label{bc2}
\sup_{\phi\in\mathcal{F},\langle\phi|\phi\rangle=1}
\langle\phi|[\hat{a}_1,e][\hat{a}_1,e]^\dag+[\hat{a}_2,e]^\dag[\hat{a}_2,e]|\phi\rangle
\leqslant\frac{\hbar}{2}.
\end{equation}

From (\ref{d1d}), we also have
\begin{equation}
D_1 D_1^\dag
=\left(
  \begin{array}{cc}
  [\hat{a}_2,e]^\dag[\hat{a}_2,e]+[\hat{a}_1,e]^\dag[\hat{a}_1,e] & [\hat{a}_2,e]^\dag[\hat{a}_1,e]^\dag-[\hat{a}_1,e]^\dag[\hat{a}_2,e]^\dag \\{}
  [\hat{a}_1,e][\hat{a}_2,e]-[\hat{a}_2,e][\hat{a}_1,e] & [\hat{a}_1,e][\hat{a}_1,e]^\dag+[\hat{a}_2,e][\hat{a}_2,e]^\dag \\
  \end{array}\right).~~
\end{equation}
Similarly, there are also the following inequalities
\begin{eqnarray}\label{bc3}
&&\sup_{\phi\in\mathcal{F},\langle\phi|\phi\rangle=1}
\langle\phi|[\hat{a}_1,e]^\dag[\hat{a}_1,e]+[\hat{a}_2,e]^\dag[\hat{a}_2,e]|\phi\rangle
\leqslant\frac{\hbar}{2},
\nonumber\\
&&\sup_{\phi\in\mathcal{F},\langle\phi|\phi\rangle=1}
\langle\phi|[\hat{a}_1,e][\hat{a}_1,e]^\dag+[\hat{a}_2,e][\hat{a}_2,e]^\dag|\phi\rangle
\leqslant\frac{\hbar}{2}.
\end{eqnarray}

\section{Optimal element and constraint relations}\label{sec4}
Consider the $2D$ harmonic oscillator states $|p,q\rangle$ and $|m,n\rangle$, and the corresponding density matrices are $\rho_{p,q}=|p,q\rangle \langle p,q|$ and $\rho_{m,n}=|m,n\rangle \langle m,n|$. The Connes distance between these two states is
\begin{eqnarray}\label{d0}
d(\omega_{p,q},\omega_{m,n})
&=&\sup_{e\in B}|\mathrm{tr}_{\mathcal{F}}(\rho_{p,q} e)-\mathrm{tr}_{\mathcal{F}}(\rho_{m,n} e)|\nonumber\\
&=&\sup_{e\in B}|\mathrm{tr}_{\mathcal{F}}(\Delta\rho\,e)|\nonumber\\
&=&\sup_{e\in B}|\langle p,q|e|p,q\rangle-\langle m,n|e|m,n\rangle|,
\end{eqnarray}
where $\Delta\rho=\rho_{p,q}-\rho_{m,n}=|p,q\rangle\langle p,q|-|m,n\rangle\langle m,n|$.
The element $e\in B$ can be expressed as
\begin{equation}
e=\sum_{i,j,k,l=0}^\infty C^{i,j}_{k,l}|i,j\rangle\langle k,l|.
\end{equation}
Consider the element $e$ attaining the supremum in (\ref{d0}), which is the so-called optimal element.
Since only Hermitian elements can give the supremum in the Connes
distance function \cite{Iochum}, there should be $C^{i,j}_{k,l}=(C^{k,l}_{i,j})^*$, and $C^{i,j}_{i,j}$ are real numbers.
After some straightforward calculations, one can obtain
\begin{equation}
d(\omega_{p,q},\omega_{m,n})=\sup_{e\in B}|C^{p,q}_{p,q}-C^{m,n}_{m,n}|.
\end{equation}

For the element $e$ (\ref{eo0}), using the relations (\ref{a1a2}), we have
\begin{eqnarray}
[\hat{a}_1,e]&=&\hat{a}_1 e-e \hat{a}_1\nonumber\\
&=&\sum_{i,j,k,l=0}^\infty C^{i,j}_{k,l}\sqrt{i}|i-1,j\rangle\langle k,l|
-\sum_{i,j,k,l=0}^\infty C^{i,j}_{k,l}|i,j\rangle\langle k+1,l|\sqrt{k+1}\nonumber\\
&=&\sum_{i,j,k,l=0}^\infty\left(C^{i+1,j}_{k,l}\sqrt{i+1}
-C^{i,j}_{k-1,l}\sqrt{k}\right)|i,j\rangle\langle k,l|,
\end{eqnarray}
and
\begin{eqnarray}
[\hat{a}_1,e][\hat{a}_1,e]^\dag&=&\left(\sum_{i,j,k,l=0}^\infty\left(C^{i+1,j}_{k,l}\sqrt{i+1}
-C^{i,j}_{k-1,l}\sqrt{k}\right)|i,j\rangle\langle k,l|\right)\nonumber\\
&&
~~\times\left(\sum_{i,j,k,l=0}^\infty\left(C^{i+1,j}_{k,l}\sqrt{i+1}
-C^{i,j}_{k-1,l}\sqrt{k}\right)^*|k,l\rangle\langle i,j|\right)\nonumber\\
&=&
\sum_{i,j=0}^\infty\left(\sum_{k,l=0}^\infty\left|C^{i+1,j}_{k,l}\sqrt{i+1}
-C^{i,j}_{k-1,l}\sqrt{k}\right|^2\right)|i,j\rangle\langle i,j|+n.d.\,,~~~
\end{eqnarray}
where ``$n.d.$'' denotes the sum of terms $|k,l\rangle\langle i,j|$ with $k\neq i$ and/or $l\neq j$, and these terms will not affect the calculations in the following content.

We also have
\begin{equation}
[\hat{a}_2,e]=\sum_{i,j,k,l=0}^\infty\left(C^{i,j+1}_{k,l}\sqrt{j+1}
-C^{i,j}_{k,l-1}\sqrt{l}\right)|i,j\rangle\langle k,l|,
\end{equation}
and
\begin{equation}
[\hat{a}_2,e]^\dag[\hat{a}_2,e]
=\sum_{i,j=0}^\infty\left(\sum_{k,l=0}^\infty\left|C^{i,j}_{k,l+1}\sqrt{l+1}
-C^{i,j-1}_{k,l}\sqrt{j}\right|^2\right)|i,j\rangle\langle i,j|+n.d.\,.
\end{equation}
So there is
\begin{eqnarray}\label{ae1}
\lefteqn{[\hat{a}_1,e][\hat{a}_1,e]^\dag+[\hat{a}_2,e]^\dag[\hat{a}_2,e]
=\sum_{i,j=0}^\infty\sum_{k,l=0}^\infty\bigg(\left|C^{i+1,j}_{k,l}\sqrt{i+1}
-C^{i,j}_{k-1,l}\sqrt{k}\right|^2}\nonumber\\
&&\qquad\qquad\qquad\qquad
+\left|C^{i,j}_{k,l+1}\sqrt{l+1}
-C^{i,j-1}_{k,l}\sqrt{j}\right|^2\bigg)|i,j\rangle\langle i,j|+n.d.\,.
\end{eqnarray}
Similarly, we also have
\begin{eqnarray}\label{ae2}
\lefteqn{[\hat{a}_1,e]^\dag[\hat{a}_1,e]+[\hat{a}_2,e][\hat{a}_2,e]^\dag
=\sum_{i,j=0}^\infty\sum_{k,l=0}^\infty\bigg(\left|C^{i,j}_{k+1,l}\sqrt{k+1}
-C^{i-1,j}_{k,l}\sqrt{i}\right|^2}\nonumber\\
&&\qquad\qquad\qquad\qquad+\left|C^{i,j+1}_{k,l}\sqrt{j+1}
-C^{i,j}_{k,l-1}\sqrt{l}\right|^2\bigg)|i,j\rangle\langle i,j|+n.d.\,.
\end{eqnarray}

From (\ref{bc1}), (\ref{bc2}) and (\ref{ae1}), (\ref{ae2}), one can obtain the following inequalities,
\begin{eqnarray}\label{cc0}
&&\sum_{k,l=0}^\infty\left(\left|C^{i,j}_{k+1,l}\sqrt{k+1}
-C^{i-1,j}_{k,l}\sqrt{i}\right|^2+\left|C^{i,j+1}_{k,l}\sqrt{j+1}
-C^{i,j}_{k,l-1}\sqrt{l}\right|^2\right)\leqslant\frac{\hbar}{2},\nonumber\\
&&\sum_{k,l=0}^\infty\left(\left|C^{i+1,j}_{k,l}\sqrt{i+1}
-C^{i,j}_{k-1,l}\sqrt{k}\right|^2+\left|C^{i,j}_{k,l+1}\sqrt{l+1}
-C^{i,j-1}_{k,l}\sqrt{j}\right|^2\right)\leqslant\frac{\hbar}{2}.~~~~~
\end{eqnarray}
For simplicity, we can set $C^{i,j}_{k,l}=0$ if $k\neq i$ and/or $l\neq j$, and denote $C^{i,j}_{i,j}=c_{i,j}$.
For $i,j=0,1,2,...$, one can obtain the following inequalities,
\begin{eqnarray}
&&\left|c_{i,j}\sqrt{i}
-c_{i-1,j}\sqrt{i}\right|^2+\left|c_{i,j+1}\sqrt{j+1}
-c_{i,j}\sqrt{j+1}\right|^2\leqslant\frac{\hbar}{2},\nonumber\\
&&\left|c_{i+1,j}\sqrt{i+1}
-c_{i,j}\sqrt{i+1}\right|^2+\left|c_{i,j}\sqrt{j}
-c_{i,j-1}\sqrt{j}\right|^2\leqslant\frac{\hbar}{2}.
\end{eqnarray}
Since $c_{i,j}=C^{i,j}_{i,j}$ are real, so we have
\begin{eqnarray}\label{cc1}
&&i\left(c_{i,j}-c_{i-1,j}\right)^2
+(j+1)\left(c_{i,j+1}-c_{i,j}\right)^2
\leqslant\frac{\hbar}{2},\nonumber\\
&&(i+1)\left(c_{i+1,j}-c_{i,j}\right)^2
+j\left(c_{i,j}
-c_{i,j-1}\right)^2\leqslant\frac{\hbar}{2}.
\end{eqnarray}
Denote
\begin{equation}
c_{i,j}-c_{i-1,j}=E_{i,j},\qquad
c_{i,j}-c_{i,j-1}=F_{i,j},
\end{equation}
then the inequalities (\ref{cc1}) can be rewritten as
\begin{equation}\label{ee1}
iE_{i,j}^2+(j+1)F_{i,j+1}^2\leqslant\frac{\hbar}{2},\quad
(i+1)E_{i+1,j}^2+jF_{i,j}^2\leqslant\frac{\hbar}{2}.
\end{equation}

Similarly, from (\ref{bc3}), one can obtain the following inequalities,
\begin{eqnarray}\label{cc2}
&&i\left(c_{i,j}-c_{i-1,j}\right)^2
+j\left(c_{i,j}-c_{i,j-1}\right)^2
\leqslant\frac{\hbar}{2},\nonumber\\
&&(i+1)\left(c_{i+1,j}
-c_{i,j}\right)^2+
(j+1)\left(c_{i,j+1}
-c_{i,j}\right)^2\leqslant\frac{\hbar}{2},
\end{eqnarray}
or
\begin{equation}\label{ee2}
iE_{i,j}^2+jF_{i,j}^2\leqslant\frac{\hbar}{2},\quad
(i+1)E_{i+1,j}^2+(j+1)F_{i,j+1}^2\leqslant\frac{\hbar}{2}.
\end{equation}

So in order to obtain the supremum in (\ref{d0}), in general, one only need to consider the so-called optimal elements $e_o\in B$ which can be expressed as
\begin{equation}\label{eo0}
e_o=\sum_{i,j=0}^\infty c_{i,j}|i,j\rangle\langle i,j|,
\end{equation}
where the coefficients $c_{i,j}$ satisfy the constraint relations (\ref{cc1}), (\ref{cc2}) or (\ref{ee1}), (\ref{ee2}), and the Connes distance (\ref{d0}) is obtained by
\begin{eqnarray}
d(\omega_{p,q},\omega_{m,n})
&=&|\mathrm{tr}_{\mathcal{F}}(\rho_{p,q} e_o)-\mathrm{tr}_{\mathcal{F}}(\rho_{m,n} e_o)|\nonumber\\
&=&|\langle p,q|e_o|p,q\rangle-\langle m,n|e_o|m,n\rangle|.
\end{eqnarray}

\section{Connes distance between harmonic oscillator states}\label{sec5}
Now let us explicitly calculate the Connes distances between the quantum states of $2D$ harmonic oscillators.
First, let us consider the adjacent states $|m+1,n\rangle$ and $|m,n\rangle$, and the corresponding Connes distance is
\begin{eqnarray}
d(\omega_{m+1,n},\omega_{m,n})&=&\sup_{e\in B}|\mathrm{tr}_{\mathcal{F}}(\rho_{m+1,n} e)-\mathrm{tr}_{\mathcal{F}}(\rho_{m,n} e)|\nonumber\\
&=&\sup_{e\in B}|c_{m+1,n}-c_{m,n}|
=\sup_{e\in B}|E_{m+1,n}|.
\end{eqnarray}

From (\ref{ee1}) and (\ref{ee2}), we have
\begin{eqnarray}
&&(m+1)E_{m+1,n}^2+(n+1)F_{m+1,n+1}^2\leqslant\frac{\hbar}{2},\quad
(m+1)E_{m+1,n}^2+nF_{m,n}^2\leqslant\frac{\hbar}{2},\nonumber\\
&&(m+1)E_{m+1,n}^2+(n+1)F_{m,n+1}^2\leqslant\frac{\hbar}{2},\quad
(m+1)E_{m+1,n}^2+nF_{m+1,n}^2\leqslant\frac{\hbar}{2}.\quad
\end{eqnarray}
So in order to attain the supremum of $|c_{m+1,n}-c_{m,n}|=|E_{m+1,n}|$, there should be
\begin{equation}\label{me}
(m+1)E_{m+1,n}^2=\frac{\hbar}{2},
\end{equation}
or
\begin{equation}
|c_{m+1,n}-c_{m,n}|=|E_{m+1,n}|=\sqrt{\frac{\hbar}{2(m+1)}},
\end{equation}
and then there should be
\begin{equation}
F_{m,n+1}=F_{m+1,n+1}=0.
\end{equation}
Without loss of generality, we can assume $E_{m+1,n}\geqslant 0$, and
\begin{eqnarray}
E_{m+1,n}&=&F_{m+1,n+1}+E_{m+1,n}=c_{m+1,n+1}-c_{m,n}\nonumber\\
&=&E_{m+1,n+1}+F_{m,n+1}=E_{m+1,n+1}.
\end{eqnarray}
So it is easy to see that, if $E_{m+1,n}$ satisfy (\ref{me}), there must be
\begin{equation}
E_{m+1,i}=\sqrt{\frac{\hbar}{2(m+1)}},\qquad
F_{m,i+1}=F_{m+1,i+1}=0,
\end{equation}
where $i=0,1,2,...$\,. This means that
\begin{equation}
c_{m,0}=c_{m,1}=...=c_{m,i}=...\,,\qquad
c_{m+1,0}=c_{m+1,1}=...=c_{m+1,i}=...\,.
\end{equation}

Furthermore, let us consider the states $|m+k,n\rangle$ and $|m,n\rangle$, $k>1$, and the corresponding Connes distance is
\begin{eqnarray}\label{d1}
d(\omega_{m+k,n},\omega_{m,n})&=&\sup_{e\in B}|\mathrm{tr}_{\mathcal{F}}(\rho_{m+k,n} e)-\mathrm{tr}_{\mathcal{F}}(\rho_{m,n} e)|
=\sup_{e\in B}|c_{m+k,n}-c_{m,n}|\nonumber\\
&=&\sup_{e\in B}|c_{m+k,n}-c_{m+k-1,n}+c_{m+k-1,n}-c_{m+k-2,n}\nonumber\\
&&\qquad+...+c_{m+1,n}-c_{m,n}|\nonumber\\
&\leqslant&\sup_{e\in B}\big(|c_{m+k,n}-c_{m+k-1,n}|
+|c_{m+k-1,n}-c_{m+k-2,n}|\nonumber\\
&&\qquad+...+|c_{m+1,n}-c_{m,n}|\big)\nonumber\\
&\leqslant&\sqrt{\frac{\hbar}{2}}
\left(\frac{1}{\sqrt{m+k}}+\frac{1}{\sqrt{m+k-1}}+...+\frac{1}{\sqrt{m+1}}\right)\nonumber\\
&=&\sqrt{\frac{\hbar}{2}}
\sum_{i=1}^{k}\frac{1}{\sqrt{m+i}}=\zeta_{m;m+k}\sqrt{\frac{\hbar}{2}},
\end{eqnarray}
where
\begin{equation}
\zeta_{p;q}=\sum_{i=p+1}^{q}\frac{1}{\sqrt{i}}
=\zeta\left(\frac{1}{2},p+1\right)
-\zeta\left(\frac{1}{2},q+1\right),
\end{equation}
and $\zeta(s,q)$ is the Hurwitz zeta function
\begin{equation}
\zeta(s,q)=\sum_{i=0}^{\infty}\frac{1}{(q+i)^s}.
\end{equation}
Obviously, there is $\zeta_{i;j}=-\zeta_{j;i}$ and $\zeta_{i;i}=0$.

In order to attain the supremum of $|c_{m+k,n}-c_{m,n}|$, for example, one can choose
\begin{eqnarray}
&&c_{0,i}=0,\nonumber\\
&&c_{1,i}
=\sqrt{\frac{\hbar}{2}},\nonumber\\
&&c_{2,i}
=\sqrt{\frac{\hbar}{2}}\left(1+\frac{1}{\sqrt{2}}\right),\nonumber\\
&&\qquad\vdots \nonumber\\
&&c_{p,i}
=\sqrt{\frac{\hbar}{2}}
\sum_{j=1}^{p}\frac{1}{\sqrt{j}}=\zeta_{0;p}\sqrt{\frac{\hbar}{2}},\nonumber\\
&&\qquad\vdots
\end{eqnarray}
where $p,i=0,1,2...$\,,
and there is
\begin{equation}
c_{m+i,n}-c_{m+i-1,n}=E_{m+i,n}
=\sqrt{\frac{\hbar}{2(m+i)}}\qquad(i=1,2,...,k).
\end{equation}

Suppose $e_o$ to be the optimal element which attains the supremum in (\ref{d1}),
it can be expressed as
\begin{equation}\label{eo11}
e_o=\sum_{p,i=0}^\infty
c_{p,i}|p,i\rangle\langle p,i|
=\sqrt{\frac{\hbar}{2}}\sum_{p,i=0}^\infty\zeta_{0;p}
|p,i\rangle\langle p,i|,
\end{equation}
and the Connes distance between the states $|m+k,n\rangle$ and $|m,n\rangle$ is
\begin{eqnarray}\label{d11}
d(\omega_{m+k,n},\omega_{m,n})&=&|\mathrm{tr}_{\mathcal{F}}(\rho_{m+k,n} e_o)-\mathrm{tr}_{\mathcal{F}}(\rho_{m,n} e_o)|
=|c_{m+k,n}-c_{m,n}|\nonumber\\
&=&\sqrt{\frac{\hbar}{2}}
\sum_{i=1}^{k}\frac{1}{\sqrt{m+i}}=\zeta_{m;m+k}\sqrt{\frac{\hbar}{2}}\nonumber\\
&=&\sqrt{\frac{\hbar}{2}}\left(\zeta\left(\frac{1}{2},m+1\right)
-\zeta\left(\frac{1}{2},m+k+1\right)\right).
\end{eqnarray}
Of course, there is $d(\omega_{m+k,n},\omega_{m,n})=d(\omega_{m,n},\omega_{m+k,n})$. These distances satisfy
\begin{equation}
d(\omega_{m+k+l,n},\omega_{m,n})
=d(\omega_{m+k,n},\omega_{m,n})+d(\omega_{m+k+l,n},\omega_{m+k,n}).
\end{equation}

These results coincide with the results of the Connes distance between the quantum states of $1D$ harmonic oscillators in Refs.~\cite{Cagnache,Scholtz,Revisiting}.
This means that the Connes spectral distance between the states $|m,n\rangle$ and $|m+k,n\rangle$ of $2D$ harmonic oscillators is the same as that between the states $|m\rangle$ and $|m+k\rangle$ of $1D$ harmonic oscillators.

Similarly, there is
\begin{eqnarray}\label{d12}
d(\omega_{m,n+k},\omega_{m,n})
&=&\sqrt{\frac{\hbar}{2}}
\sum_{i=1}^{k}\frac{1}{\sqrt{n+i}}
=\zeta_{n;n+k}\sqrt{\frac{\hbar}{2}}\nonumber\\
&=&\sqrt{\frac{\hbar}{2}}\left(\zeta\left(\frac{1}{2},n+1\right)
-\zeta\left(\frac{1}{2},n+k+1\right)\right).
\end{eqnarray}

\section{Distance between arbitrary Fock states $|m,n\rangle$ and $|k,l\rangle$}\label{sec6}
Now, let us consider the Connes distance between two arbitrary Fock states $|m,n\rangle$ and $|k,l\rangle$.
First, let us consider the states $|m+1,n+1\rangle$ and $|m,n\rangle$, and the corresponding Connes distance is
\begin{eqnarray}
\lefteqn{d(\omega_{m+1,n+1},\omega_{m,n})
=\sup_{e\in B}|\mathrm{tr}_{\mathcal{F}}(\rho_{m+1,n+1} e)-\mathrm{tr}_{\mathcal{F}}(\rho_{m,n} e)|}\nonumber\\
&=&\sup_{e\in B}|c_{m+1,n+1}-c_{m,n}|
=\sup_{e\in B}|E_{m+1,n+1}+F_{m,n+1}|\nonumber\\
&=&\sup_{e\in B}\left|\frac{1}{\sqrt{m+1}}E_{m+1,n+1}\sqrt{m+1}
+\frac{1}{\sqrt{n+1}}F_{m,n+1}\sqrt{n+1}\right|\nonumber\\
&\leqslant&\sup_{e\in B}\sqrt{\frac{1}{m+1}+\frac{1}{n+1}}
\sqrt{(m+1)E_{m+1,n+1}^2+(n+1)F_{m,n+1}^2}\nonumber\\
&\leqslant&\sqrt{\frac{1}{m+1}+\frac{1}{n+1}}\sqrt{\frac{\hbar}{2}}.
\end{eqnarray}
In the first inequality, we have used the Cauchy-Schwartz inequality,
\begin{equation}\label{cs}
(x_1y_1+x_2y_2)^2\leqslant(x_1^2+x_2^2)(y_1^2+y_2^2),
\end{equation}
where equality holds if $x_1 y_2=x_2 y_1$.

From (\ref{ee1}), in order to attain the supremum of $|E_{m+1,n+1}+F_{m,n+1}|$, there should be
\begin{equation}\label{ef1}
(m+1)E_{m+1,n+1}^2+(n+1)F_{m,n+1}^2=\frac{\hbar}{2},
\end{equation}
and
\begin{equation}
(m+1)E_{m+1,n+1}=(n+1)F_{m,n+1}.
\end{equation}
Without loss of generality, we can assume that $E_{m+1,n+1},F_{m,n+1}\geqslant0$. From (\ref{ee2}), we have
\begin{eqnarray}
&&(m+1)E_{m+1,n+1}^2+(n+1)F_{m+1,n+1}^2\leqslant\frac{\hbar}{2},\nonumber\\
&&(m+1)E_{m+1,n}^2+(n+1)F_{m,n+1}^2\leqslant\frac{\hbar}{2}.
\end{eqnarray}
Compare the above inequalities with (\ref{ef1}), one can obtain
\begin{equation}
E_{m+1,n}\leqslant E_{m+1,n+1},\qquad
F_{m+1,n+1}\leqslant F_{m,n+1}.
\end{equation}
But
\begin{equation}
E_{m+1,n}+F_{m+1,n+1}=E_{m+1,n+1}+F_{m,n+1}=c_{m+1,n+1}-c_{m,n},
\end{equation}
so there must be
\begin{equation}
E_{m+1,n}=E_{m+1,n+1},\qquad
F_{m+1,n+1}=F_{m,n+1}.
\end{equation}

Furthermore, let us consider the Fock states $|m+2,n+1\rangle$ and $|m,n\rangle$, and the corresponding Connes distance is
\begin{eqnarray}
d(\omega_{m+2,n+1},\omega_{m,n})
&=&\sup_{e\in B}|\mathrm{tr}_{\mathcal{F}}(\rho_{m+2,n+1} e)-\mathrm{tr}_{\mathcal{F}}(\rho_{m,n} e)|
=\sup_{e\in B}|c_{m+2,n+1}-c_{m,n}|\nonumber\\
&=&\sup_{e\in B}|E_{m+2,n+1}+F_{m+1,n+1}+E_{m+1,n}|.
\end{eqnarray}
In order to attain the supremum of $|E_{m+2,n+1}+F_{m+1,n+1}+E_{m+1,n}|$, there must be
\begin{eqnarray}\label{efef}
&&(m+1)E_{m+1,n}^2+(n+1)F_{m+1,n+1}^2=\frac{\hbar}{2},\nonumber\\
&&(m+2)E_{m+2,n+1}^2+(n+1)F_{m+1,n+1}^2=\frac{\hbar}{2},
\end{eqnarray}
and then
\begin{equation}
(m+1)E_{m+1,n}^2=(m+2)E_{m+2,n+1}^2.
\end{equation}
Without loss of generality, we can assume $E_{m+2,n+1}, E_{m+1,n}, F_{m+1,n+1}\geqslant 0$.
It is easy to see that, there must be
\begin{equation}
E_{m+1,n+1}=E_{m+1,n}=\sqrt{\frac{m+2}{m+1}}E_{m+2,n+1}=\sqrt{\frac{m+2}{m+1}}E_{m+2,n},
\end{equation}
\begin{equation}
F_{m+2,n+1}=F_{m+1,n+1}=F_{m,n+1}.
\end{equation}
So we have
\begin{eqnarray}
\lefteqn{d(\omega_{m+2,n+1},\omega_{m,n})
=\sup_{e\in B}|E_{m+2,n+1}+F_{m+1,n+1}+E_{m+1,n}|}\nonumber\\
&=&\sup_{e\in B}\left|\left(\frac{1}{\sqrt{m+1}}+\frac{1}{\sqrt{m+2}}\right)E_{m+1,n}\sqrt{m+1}
+\frac{1}{\sqrt{n+1}}F_{m+1,n+1}\sqrt{n+1}\right|\nonumber\\
&\leqslant&\sup_{e\in B}\sqrt{(m+1)E_{m+1,n}^2+(n+1)F_{m+1,n+1}^2}
\sqrt{\left(\frac{1}{\sqrt{m+1}}+\frac{1}{\sqrt{m+2}}\right)^2+\frac{1}{n+1}}
\nonumber\\
&\leqslant&\sqrt{\frac{\hbar}{2}}
\sqrt{\left(\frac{1}{\sqrt{m+1}}+\frac{1}{\sqrt{m+2}}\right)^2+\frac{1}{n+1}}.
\end{eqnarray}
Here in the first inequality, we have used the Cauchy-Schwartz inequality (\ref{cs}).

Finally, let us consider the Connes distance between two arbitrary Fock states $|m,n\rangle$ and $|m+k,n+l\rangle$. For simplicity, we assume $k,l\geqslant 0$. The corresponding Connes distance is
\begin{eqnarray}
\lefteqn{d(\omega_{m+k,n+l},\omega_{m,n})
=\sup_{e\in B}|\mathrm{tr}_{\mathcal{F}}(\rho_{m+k,n+l} e)-\mathrm{tr}_{\mathcal{F}}(\rho_{m,n} e)|}\nonumber\\
&=&\sup_{e\in B}|c_{m+k,n+l}-c_{m,n}|
=\sup_{e\in B}|c_{m+k,n+l}-c_{m+k,n}+c_{m+k,n}-c_{m,n}|\nonumber\\
&=&\sup_{e\in B}\left|\sum_{i=1}^{l}(c_{m+k,n+i}-c_{m+k,n+i-1})
+\sum_{j=1}^{k}(c_{m+j,n}-c_{m+j-1,n})\right|\nonumber\\
&=&\sup_{e\in B}\left|\sum_{i=1}^{l}F_{m+k,n+i}+\sum_{j=1}^{k}E_{m+j,n}\right|.
\end{eqnarray}
Similarly, in order to attain the supremum of $|c_{m+k,n+l}-c_{m,n}|$,
there must be
\begin{equation}\label{me1}
(m+1)E_{m+1,n}^2+(n+1)F_{m,n+1}^2=\frac{\hbar}{2},
\end{equation}
and
\begin{equation}
E_{m+i,n+j}=\sqrt{\frac{m+1}{m+i}}E_{m+1,n}
\qquad(1\leqslant i\leqslant k,0\leqslant j\leqslant l),
\end{equation}
\begin{equation}
F_{m+i,n+j}=\sqrt{\frac{n+1}{n+j}}F_{m,n+1}
\qquad(0\leqslant i\leqslant k,1\leqslant j\leqslant l).
\end{equation}
So we have
\begin{eqnarray}
\lefteqn{d(\omega_{m+k,n+l},\omega_{m,n})
=\sup_{e\in B}\left|\sum_{i=1}^{k}E_{m+i,n}+\sum_{j=1}^{l}F_{m+k,n+j}\right|}\nonumber\\
&=&\sup_{e\in B}\left|\sqrt{m+1}E_{m+1,n}\sum_{i=1}^{k}\frac{1}{\sqrt{m+i}}
+\sqrt{n+1}F_{m,n+1}\sum_{j=1}^{l}\frac{1}{\sqrt{n+j}}\right|\nonumber\\
&=&\sup_{e\in B}\left|\sqrt{m+1}E_{m+1,n}\zeta_{m;m+k}
+\sqrt{n+1}F_{m,n+1}\zeta_{n;n+l}\right|\nonumber\\
&\leqslant&\sup_{e\in B}
\sqrt{(m+1)E_{m+1,n}^2+(n+1)F_{m,n+1}^2}
\sqrt{\zeta_{m;m+k}^2+\zeta_{n;n+l}^2}\nonumber\\
&=&\sqrt{\frac{\hbar}{2}}
\sqrt{\zeta_{m;m+k}^2+\zeta_{n;n+l}^2}.
\end{eqnarray}
Here we have used the Cauchy-Schwartz inequality (\ref{cs}) in the above inequality, and the equality holds if
\begin{equation}\label{ef}
\sqrt{m+1}E_{m+1,n}\zeta_{n;n+l}=\sqrt{n+1}F_{m,n+1}\zeta_{m;m+k}.
\end{equation}
From (\ref{me1}) and (\ref{ef}), one can derive
\begin{equation}
E_{m+1,n}=\sqrt{\frac{\hbar}{2(m+1)}}
\frac{\zeta_{m;m+k}}{\sqrt{\zeta_{m;m+k}^2
+\zeta_{n;n+l}^2}}=c_0\frac{1}{\sqrt{m+1}},
\end{equation}
\begin{equation}
F_{m,n+1}=\sqrt{\frac{\hbar}{2(n+1)}}
\frac{\zeta_{n;n+l}}{\sqrt{\zeta_{m;m+k}^2
+\zeta_{n;n+l}^2}}=d_0\frac{1}{\sqrt{n+1}},
\end{equation}
where
\begin{equation}
c_0=\sqrt{\frac{\hbar}{2}}
\frac{\zeta_{m;m+k}}{\sqrt{\zeta_{m;m+k}^2
+\zeta_{n;n+l}^2}},\qquad
d_0=\sqrt{\frac{\hbar}{2}}
\frac{\zeta_{n;n+l}}{\sqrt{\zeta_{m;m+k}^2
+\zeta_{n;n+l}^2}}.
\end{equation}

So in order to attain the supremum of $|c_{m+k,n+l}-c_{m,n}|$, for example, one can choose
\begin{eqnarray}
&&c_{0,0}=0,\quad c_{1,0}=c_0,\quad c_{0,1}=d_0,\quad c_{1,1}=c_0+d_0,\quad ...\nonumber\\
&&c_{p,q}=c_0\sum_{i=1}^{p}\frac{1}{\sqrt{i}}
+d_0\sum_{j=1}^{q}\frac{1}{\sqrt{j}}
=c_0\zeta_{0;p}+d_0\zeta_{0;q}\nonumber\\
&&~~~~=\sqrt{\frac{\hbar}{2}}
\frac{\zeta_{0;p}\zeta_{m;m+k}+\zeta_{0;q}\zeta_{n;n+l}}{\sqrt{\zeta_{m;m+k}^2
+\zeta_{n;n+l}^2}},\quad ...
\end{eqnarray}
where $p,q=0,1,2,...$\,.
Obviously, there is
\begin{equation}
E_{i+1,j}=c_{i+1,j}-c_{i,j}=\frac{1}{\sqrt{i+1}}c_0,
\qquad
F_{i,j+1}=c_{i,j+1}-c_{i,j}=\frac{1}{\sqrt{j+1}}d_0.
\end{equation}
The corresponding optimal element $e_o$ is
\begin{equation}\label{eo1}
e_o=\sum_{p,q=0}^\infty
c_{p,q}|p,q\rangle\langle p,q|
=\sqrt{\frac{\hbar}{2}}\sum_{p,q=0}^\infty
\frac{\zeta_{0;p}\zeta_{m;m+k}+\zeta_{0;q}\zeta_{n;n+l}}{\sqrt{\zeta_{m;m+k}^2
+\zeta_{n;n+l}^2}}|p,q\rangle\langle p,q|,
\end{equation}
and the Connes distance between the states $|m+k,n+l\rangle$ and $|m,n\rangle$ is
\begin{eqnarray}\label{d2}
d(\omega_{m+k,n+l},\omega_{m,n})
&=&|\mathrm{tr}_{\mathcal{F}}(\rho_{m+k,n+l} e_o)-\mathrm{tr}_{\mathcal{F}}(\rho_{m,n} e_o)|
=|c_{m+k,n+l}-c_{m,n}|\nonumber\\
&=&c_0\zeta_{m;m+k}+d_0\zeta_{n;n+l}\nonumber\\
&=&\sqrt{\frac{\hbar}{2}}
\sqrt{\zeta_{m;m+k}^2+\zeta_{n;n+l}^2}.
\end{eqnarray}

Obviously, when $l=0$, (\ref{eo1}) and (\ref{d2}) will return to (\ref{eo11}) and (\ref{d11}), respectively.
It is easy to verify that, these results (\ref{eo1}) and (\ref{d2}) are also true for $k<0$ and/or $l<0$.
Furthermore, we also have
\begin{eqnarray}
d(\omega_{m+k,n+l},\omega_{m,n})
&=&d(\omega_{m,n+l},\omega_{m+k,n})
=d(\omega_{m+k,n},\omega_{m,n+l})\nonumber\\
&=&d(\omega_{m,n},\omega_{m+k,n+l}).
\end{eqnarray}

From (\ref{d11}), (\ref{d12}) and (\ref{d2}), there is
\begin{eqnarray}
d(\omega_{m+k,n+l},\omega_{m,n})
&=&\sqrt{d(\omega_{m+k,n+l},\omega_{m+k,n})^2+d(\omega_{m+k,n},\omega_{m,n})^2}\nonumber\\
&=&\sqrt{d(\omega_{m+k,n+l},\omega_{m,n+l})^2+d(\omega_{m,n+l},\omega_{m,n})^2}.
\end{eqnarray}
So these distances also satisfy the Pythagoras theorem.
This also coincides with the result in Ref.~\cite{DAndrea}.

Furthermore, one can also analyse the Connes distance between mixed states. But usually, the calculations and expressions are much more cumbersome.

\section{Discussions and conclusions}\label{sec7}
In this paper, we study the Connes distance of quantum states of $2D$ harmonic oscillators in phase space. By virtue of the Hilbert-Schmidt operatorial formulation, we construct a boson Fock space and a quantum Hilbert space, and obtain the Dirac operator and a spectral triple corresponding to a $4D$ quantum phase space.
From the ball condition, we obtain some constraint relations about the optimal elements.
Using these constraint relations, we derive the explicit expressions of the Connes distance between the states of $2D$ quantum harmonic oscillators.

We find that the Connes distance between two arbitrary Fock states is not just a simple sum of distances of adjacent Fock states.
Furthermore, these distances also satisfy the Pythagoras theorem.
The calculations and results of the Connes distance of quantum states of $2D$ harmonic oscillators are much more complicated than those of the states of $1D$ harmonic oscillators.
These results are not just trivial generalizations of the result of $1D$ case.
Our method used in the present work is also different from those used in the literatures.
So these results are significant for the study of the Connes distances of physical systems in noncommutative spaces.

Here we only analyse the Connes distance between two pure states.
One can also study the distance between two mixed states.
Furthermore, one can use our method to study the Connes distance between the quantum states of higher-dimensional harmonic oscillators.
But usually the calculations and results will be much more complicated.

We are still exploring the intuitive physical meaning of these distances of physical systems.
Our methods can be used to study other physical systems in other kinds of noncommutative spaces. We hope that our results can help to study the mathematical structures and physical properties of noncommutative spaces.

\section*{Acknowledgements}
This work is partly supported by the National Natural Science Foundation of China (Grant No.~11911530750), the Guangdong Basic and Applied Basic Research Foundation (Grant No.~2019A1515011703), the Fundamental Research Funds for the Central Universities (Grant No.~2019MS109), Key Research and Development Project of Guangdong Province (Grant No.~2020B0303300001) and the Natural Science Foundation of Anhui Province (Grant No.~1908085MA16).

\end{document}